\documentstyle[prl,aps]{revtex}
\begin{document}

\draft

\preprint{gr-qc/0101079}

\title{Note on Scalar Fields Non--Minimally Coupled to $(2+1)$--Gravity}

\author{Eloy Ay\'on--Beato$^{\P}$\thanks{E-mail: ayon@fis.cinvestav.mx},
Alberto Garc\'{\i}a$^{\P}$\thanks{E-mail:
aagarcia@fis.cinvestav.mx}, Alfredo
Mac\'{\i}as$^{\P\diamond}$\thanks{E-mail: amac@xanum.uam.mx;
macias@fis.cinvestav.mx}, and Jos\'e M.\
P\'erez--S\'anchez$^{\star}$\thanks{E-mail:
iosephus@ictp.trieste.it}}

\address{$^{\P}$~Departamento~de~F\'{\i}sica,~CINVESTAV--IPN,\\
Apartado~Postal~14--740,~C.P.~07000,~M\'{e}xico,~D.F.,~MEXICO.\\
$^{\diamond}$~Departamento~de~F\'{\i}sica,~UAM--Iztapalapa,\\
Apartado~Postal~55--534,~C.P.~09340,~M\'{e}xico,~D.F.,~MEXICO.\\
$^{\star}$~Diploma~Programme~HEP,~The~Abdus~Salam~ICTP,\\
P.O.~Box~586,~34100~Trieste,~ITALY.}

\date{\today }

\maketitle

\begin{abstract}
Scalar fields non--minimally coupled to $(2+1)$--gravity, in the
presence of cosmological constant term, are considered.
Non--minimal couplings are described by the term $\zeta \,R\,\Psi
^2$ in the Lagrangian. Within a class of static circularly
symmetric space--times, it is shown that the only existing
physically relevant solutions are the anti--de~Sitter space--time
for $\zeta =0$, and the Mart\'{\i}nez--Zanelli black hole for
$\zeta =1/8$. We obtain also two new solutions with non--trivial
scalar field, for $\zeta=1/6$ and $\zeta=1/8$ respectively,
nevertheless, the corresponding space--times can be reduced, via
coordinate transformations, to the standard anti--de~Sitter
space.\\ {\bf file scalar.tex, 19.10.2000 }.
\end{abstract}

\pacs{PACS numbers: 04.50.+h, 04.60.Kz, 04.20.Jb, 04.70.Bw}

Conformal solutions have been extensively studied in General
Relativity, starting from most the general action coupled to
scalar fields \cite{berg,wag,sing}
\begin{equation}
S=\frac 12\int d^D x\sqrt{-g}
\left( \frac 1\kappa C(\Psi )R-w(\Psi )\nabla
_\mu \Psi \nabla ^\mu \Psi +V(\Psi )\right) \, ,
\label{mga}
\end{equation}
where $R$ is the Ricci scalar, $V(\Psi )$ is a potential function,
$C(\Psi )$ and $w(\Psi )$ are coupling functions.

In four dimensions, the conformal solution discovered by
Bekenstein \cite{Bekens74}, appears to be the only non--trivial
black hole solution allowed for self--gravitating scalar fields
non--minimally coupled to gravity. The study of the relevant
system in $D>2+1$ dimensions has shown that some restrictions on
the kind of plausible scalar field behaviors are required
\cite{Saa96b}. For example, in $(3+1)$--dimensions, the most
general result including self--interactions and without
cosmological constant, has been obtained by Mayo and Bekenstein
\cite{MayoBekens96,Bekens96}. In that work, spherical static
non--trivial scalar field behaviors are excluded for non--minimal
couplings with $\zeta <0$ and $\zeta \geq 1/2$. Consequently,
since in $(3+1) $--dimensions the conformal coupling is $\zeta
=1/6$ the Bekenstein black hole belongs to the non--covered range
($0<\zeta <1/2$). In this sense, still remains open the question
if in $4$--dimensions the value $\zeta =1/6$ is the unique
coupling allowing a non--trivial scalar field behavior, or if
there exist a family of solutions with non--minimally coupled
scalar field behaviors within the interval under consideration.

Although the space--time is not three--dimensional, and
$(2+1)$--dimensional gravity is clearly not a physically realistic
model of our universe, it is a simple model which is rich enough
to allow us to learn a good deal about the nature of quantum
gravity. At first sight, the $(2+1)$--dimensional gravity looks
trivial, in particular, the vacuum Einstein equations imply that
space--time is locally flat, corresponding to the absence of the
Weyl tensor in three dimensions. The triviality of local geometry
in $(2+1)$--dimensional gravity holds even if the cosmological
constant term is taken into account; the Einstein space is a space
of constant curvature.

However, the black hole riddle has long been one of the most
outstanding problems of modern physics. It has remained in focus
for a long time as one of the potential testing grounds for
quantum gravitational phenomena. As it is well known, three
dimensional gravity, in vacuum, admits only the trivial locally
flat $(2+1)$--Minkowski space. Thus, it is necessary either to
couple matter to the theory, e.g., a cosmological constant or
scalar matter, or to consider alternative vacuum or non--vacuum
gravity theories in order to get solutions different from the
trivial one. New solutions in $(2+1)$--gravity coupled to a
self--interacting dilaton and in vacuum scalar--tensor theories
have been obtained by Chan \cite{chan}. Moreover, Chan and Mann
\cite{chan1} determined a conformal static black hole solution
with a nontrivial conformal factor $\Psi $ for $C(\Psi )=1$,
$w(\Psi )=4$, and $V(\Psi )=2\Lambda \exp(b\Psi )$. A black hole
solution with a negative cosmological constant coupled to a
conformal scalar field for $C(\Psi )=1-\kappa (1/8)\Psi ^2$,
$w(\Psi )=1$, and $V(\Psi)=2\Lambda/\kappa$ has been found by
Mart\'{\i}nez and Zanelli \cite{MartZan96}.

In this paper we consider a particular case of the action
(\ref{mga}) in order to study non--minimal couplings of scalar
fields to $(2+1)$--gravity with a cosmological constant term.
Essentially, we work out a commonly used one--parametric family of
theories with real parameter $\zeta$, where the non--minimal
couplings are described by the term $\zeta\,R\,\Psi ^2$ in the
Lagrangian. We will show that for a static circularly symmetric
metric, where the space--time possesses only one degree of
freedom, the only existing non--trivial solutions are the
anti--de~Sitter space--time for $\zeta=0$, the
Mart\'{\i}nez--Zanelli black hole for $\zeta=1/8$. We obtain also
two new solutions with non--trivial scalar field, for $\zeta=1/6$
and $\zeta=1/8$ respectively, whose space--time geometries reduce
to the one of an anti--de~Sitter space. In this way we are
establishing, for the studied class, the completeness of the
solutions to the corresponding field equations.

As mentioned above, in three dimensions our basic action reads
\begin{equation}
{\cal S}=\frac 12\int d^3x\sqrt{-g}\left( \frac 1\kappa \left[
R+2l^{-2}\right] -\nabla _\mu \Psi \nabla ^\mu \Psi -\zeta
\,R\,\Psi^2\right)
,  \label{eq:ac}
\end{equation}
where $\Lambda=-l^{-2}$ is the cosmological constant, $\Psi$ is
the massless non--minimally coupled scalar field, and $R$ is the
scalar curvature.

The field equations arising from (\ref{eq:ac}) are,
on one hand, the Einstein equations
\begin{equation}
G_\mu^{~\nu }=l^{-2}\delta _\mu ^{~\nu }
+\kappa \left[ \nabla _\mu \Psi
\nabla ^\nu \Psi
-\frac 12\delta _\mu ^{~\nu }\nabla _\alpha \Psi \nabla
^\alpha \Psi
+\zeta \left( \delta _\mu ^{~\nu }\Box \Psi ^2-\nabla _\mu
\nabla ^\nu \Psi ^2+G_\mu ^{~\nu }\Psi ^2\right) \right] \, ,
\label{eq:Ein}
\end{equation}
and, on the other hand, the equation
\begin{equation}
\Box \Psi =\zeta \,R\,\Psi ,  \label{eq:scalar}
\end{equation}
for the scalar field, where $\Box$ is the Laplace--Beltrami
operator.

We shall restrict our study to the following class of static
circularly symmetric three--dimensional metrics, which, in polar
coordinates, can be written as follows
\begin{equation}
\text{\boldmath$g$}=-F(r)\text{\boldmath$dt$}^2+F(r)^{-1}
\text{\boldmath$dr$}^2+r^2\text{\boldmath$d\theta $}^2
\label{eq:circular}\, ,
\end{equation}
consequently, we assume that the scalar field only depends on the
radial variable $r$, i.e., $\Psi=\Psi(r)$.

The Einstein equations (\ref{eq:Ein}), for the static circularly
symmetric space--time (\ref{eq:circular}), become
\begin{equation}
2\zeta\Psi\Psi^{\prime\prime}+(2\zeta-1)(\Psi^{\prime})^2=0\, ,
\label{eq:Ein1}
\end{equation}
\begin{equation}
\left( 1-\kappa \zeta \Psi ^2
-2\kappa \zeta r\Psi \Psi ^{\prime }\right)
F^{\prime }-\kappa \Psi ^{\prime }
\left( 4\zeta \Psi +r\Psi ^{\prime}\right)
F =\frac{2r}{l^2} \, ,  \label{eq:Ein2}
\end{equation}

\begin{equation}
\left( 1-\kappa \zeta \Psi ^2\right) F^{\prime \prime }
-4\kappa \zeta \Psi
\Psi ^{\prime }F^{\prime }
-\kappa \left( 4\zeta \Psi \Psi ^{\prime \prime}
+\left( 4\zeta -1\right)
\left( \Psi ^{\prime }\right) ^2\right) F=\frac
2{l^2}
\label{eq:Ein3}\, .
\end{equation}
where primes denote derivatives with respect to $r$. Eq.\
(\ref{eq:Ein1}) corresponds to the combination $G_r{}^r-G_t{}^t$,
and the Eqs.\ (\ref{eq:Ein2}) and (\ref{eq:Ein3}) are the
components $G_r{}^r$ and $G_\theta{}^\theta $ of the Einstein
equations (\ref{eq:Ein}), respectively. It is straightforward to
show that the scalar field equation (\ref{eq:scalar}) follows from
Eqs.\ (\ref{eq:Ein1})--(\ref{eq:Ein3}) by using the Bianchi
identity $\nabla _\nu G_\mu {}^\nu =0$.

Let us proceed to integrate the field equations. For the general
case, $\zeta\neq 0$, Eq.\ (\ref{eq:Ein1}) becomes
\begin{equation}
\left( \frac{\Psi ^{\prime }}\Psi \right) ^{\prime }
-\frac{1-4\zeta}{2\zeta} \left( \frac{\Psi ^{\prime }}\Psi \right)
^2=0 \, , \label{eq:Ein1r}
\end{equation}
whose general solution can be immediately obtained, namely
\begin{equation}
\Psi (r)=\frac{ A}{\left( r+B\right)^{2\zeta/(1-4\zeta)}} \, ,
\label{eq:solPsi}
\end{equation}
where $A$ and $B$ are integration constants.

By substituting the expression (\ref{eq:solPsi}) for $\Psi$ into
Eq.\ (\ref {eq:Ein2}), the following linear first order
differential equation for the structural function $F$ is obtained
\begin{eqnarray}
F^{\prime } &=&\frac{\kappa \delta ^2A^2\left(
\left( \delta -1\right)
r-2B\right) }{\left( r+B\right) \left( \kappa \delta A^2
\left( \left( \delta
-1\right) r-B\right) +4\left( \delta +1\right)
\left( r+B\right) ^{\delta
+1}\right) }F  \nonumber \\
&&+\frac{8\left( \delta +1\right) r
\left( r+B\right) ^{\delta +1}}{l^2\left(
\kappa \delta A^2\left( \left( \delta -1\right) r-B\right)
+4\left( \delta
+1\right) \left( r+B\right) ^{\delta +1}\right) } \, ,
\label{eq:forderF}
\end{eqnarray}
with $\delta \equiv 4\zeta /(1-4\zeta )$. It is straightforward to
show that the general solution of the Eq.\ (\ref{eq:forderF}) is
given by
\begin{equation}
F(r)=\frac{4\left( \left( \delta +1\right)
\left( r^2-B^2\right)
-l^2C\right) \left( r+B\right) ^{\delta +1}}{l^2
\left( \kappa \delta
A^2\left( \left( \delta -1\right) r-B\right)
+4\left( \delta +1\right)
\left( r+B\right) ^{\delta +1}\right) } \, ,
\label{eq:solF}
\end{equation}
with $C$ a new integration constant.

It remains to fulfill Eq.\ (\ref{eq:Ein3}), which via Eq.\
(\ref{eq:Ein1}), becomes
\begin{equation}
\left( 1-\kappa \zeta \Psi ^2\right) F^{\prime \prime }
-4\kappa \zeta \Psi
\Psi ^{\prime }F^{\prime }-\kappa
\left( \Psi ^{\prime }\right) ^2F
=\frac 2{l^2}
\label{eq:Ein3r}\, .
\end{equation}
It is easy to see that Eq.\ (\ref{eq:Ein3r}) imposes constraints
on the integration constants of the form $A=A(B,\zeta )$ and
$C=C(B,\zeta)$. Moreover, by replacing in Eq.\ (\ref{eq:Ein3r})
the expressions (\ref{eq:solPsi}) and (\ref{eq:solF}) for
$\Psi(r)$ and $F(r)$, and after some lengthy manipulations, the
following algebraic equation is obtained:
\begin{eqnarray}
8\left( \delta +1\right) ^3 &\bigg[&\left( \delta -1\right) \left(
\delta -2\right) y^4-3\delta \left( \delta -1\right) By^3-\delta
\left( l^2C-2\left( \delta +1\right) B^2\right) y^2  \nonumber \\
&&+\left( \delta +2\right) Bl^2Cy\bigg] y^{2\delta }  \nonumber \\
-2\kappa \delta \left( \delta +1\right) A^2 &\bigg[& \left( \delta
+4\right) \left( \delta +1\right) \left( \delta -1\right) ^2y^4
-\delta \left( \delta +1\right) \left( 4\delta ^2+7\delta
-9\right) By^3  \nonumber \\ &&-\delta \left( \left( \delta
-1\right) ^2l^2C -\left( \delta +1\right) \left( \delta +2\right)
\left( 5\delta +1\right) B^2\right)y^2 \nonumber
\\
&&+\left( \delta +1\right) B\left( \left( 2\delta^2-3\delta
+4\right) l^2C-2\delta \left( \delta +1\right) \left( \delta
+2\right) B^2\right) y \nonumber \\ &&-\delta ^2\left( \delta
+1\right) B^2l^2C\bigg]y^\delta  \nonumber \\ +\kappa ^2\delta
^2A^4 &\bigg[&\left( \delta +1\right) \left( \delta -1\right)
^2y^4-3\delta \left( \delta +1\right) \left( \delta -1\right)
By^3+3\delta ^2\left( \delta +1\right) B^2y^2  \nonumber \\
&&-B\left( \left( \delta -1\right) l^2C+\delta \left( \delta
+1\right) \left( \delta +2\right) B^2\right) y\bigg]=0 \, ,
\label{eq:CEq}
\end{eqnarray}
where $y$ stands for $y\equiv r+B$. Since the powers of $y$ are
linearly independent functions \cite{Elsgoltz}, the corresponding
coefficients must vanish independently. Therefore, different
possibilities arise by assigning values or ranges of values to the
parameter $\delta $:

Let us first consider the case of positive values of $\delta$. For
$\delta >0$, the highest power of $y$ in Eq.\ (\ref {eq:CEq}) is
$2\delta+4$, equating its coefficient to zero one gets the
condition
\begin{equation}
8\left( \delta +1\right) ^3\left( \delta -1\right)\left( \delta
-2\right) =0 \, .  \label{ogt}
\end{equation}
Hence, the possible solutions only exist for the values $\delta=1$
($\zeta=1/8$) and $\delta =2$ ($\zeta =1/6$) of the coupling
constant. For the first value, i.e., $\delta =1$ ($\zeta =1/8$),
Eq.\ (\ref{eq:CEq}) becomes
\begin{equation}
2\left( \kappa A^2B-4Cl^2+16B^2\right)\left( 8y^4+3\kappa
A^2By^2\right) -2B\left( 3\kappa A^2B-4Cl^2\right) \left(
24y^3+B\kappa A^2y\right) =0 \, , \label{eq:CEqd1}
\end{equation}
therefore, the vanishing of the coefficients of even and odd
powers of $y$ yields the following relations
\begin{equation}
\kappa A^2B-4Cl^2+16B^2=0=B\left( 3\kappa A^2B-4Cl^2\right)\, .
\end{equation}
Thus, in this case ($\delta=1$, $\zeta=1/8$), there exist two
admissible classes of solutions for the integration constants. The
first class is obtained for
\begin{equation}
A^2=\frac{8B}\kappa\, ,\quad C=\frac{6B^2}{l^2}\, ,
\end{equation}
and it corresponds to the black hole of Mart\'{\i}nez and Zanelli
\cite{MartZan96}, i.e.,
\begin{equation}
\Psi _{{\rm MZ}}=\sqrt{\frac{8B}{\kappa \left( r+B\right) }}\, ,
\quad F_{{\rm MZ}} =\frac{\left( r+B\right) ^2\left(
r-2B\right)}{rl^2} \label{eq:MZ}\, ,
\end{equation}
whose horizon is located at $r_{{\rm {h}}}=2B$. In this case the
free parameter $B$ is related to the mass $M$ of the hole through
the relation $B=\sqrt{Ml^2/3}$.

The second class is achieved by setting $B=C=0$, and can be
expressed as follows
\begin{equation}
\Psi (r)=\frac A{\sqrt{r}}\,,\quad F(r)=\frac{r^2}{l^2}\,.
\label{eq:d1}
\end{equation}
This is a new conformal solution with a non--trivial scalar field
behavior, although the corresponding space--time can be brought,
via coordinate transformations, to the standard anti--de~Sitter
space form (see for example
\cite{Einsenhart,Plebanski67,Weimberg,Garcia00}). Therefore, since
the anti--de~Sitter space--time is solution to the vacuum plus
cosmological constant field equations, the energy--momentum tensor
of the non--trivial scalar field is such that it vanishes
identically (see the term in brackets on the left hand side of
Eq.\ (\ref{eq:Ein})). This peculiar behavior arises from the
non--minimal coupling of the conformal scalar field to gravity.

For the second value, i.e., $\delta =2$ ($\zeta =1/6$). The
highest power of $y$ in (\ref{eq:CEq}) is $y^{2\delta +3}=y^7$,
its coefficient vanishes if the constant $B$ is set equal to zero,
then Eq.\ (\ref{eq:CEq}) reduces to
\begin{equation}
-12\left( 2Cl^2+\kappa A^2\right) \left( 18y^6-\kappa
A^2y^4\right)=0\,.
\label{eq:CEqd2}
\end{equation}
Consequently
\begin{equation}
C=-\frac{\kappa A^2}{2l^2}\,,
\end{equation}
and therefore,
\begin{equation}
\Psi (r)=\frac Ar\,,\quad F(r)=\frac{r^2}{l^2}\,.
\label{eq:d2}
\end{equation}
This is a new solution with a non--minimally coupled scalar field,
however, it exhibits the same peculiarities as the ones of the
previous solution Eq.\ (\ref{eq:d1}).

In full, for positive values of the parameter $\delta$, we have
established the existence of three classes of solutions, namely
the Mart\'{\i}nez--Zanelli black hole for $\zeta=1/8$, and two new
solutions corresponding to the anti--de~Sitter space with
non--trivial scalar fields, one for $\zeta=1/6$, and one for
$\zeta=1/8$.

Let us turn to the case of negative values of $\delta$. For
$\delta<0$ the highest power of $y$ in Eq.\ (\ref{eq:CEq}) is
$y^4$, the vanishing of its coefficient leads to the condition
\begin{equation}
\kappa ^2\delta ^2\left( \delta +1\right) \left(\delta -1\right)
^2A^4=0 \, .  \label{eq:d<0}
\end{equation}
It should be noticed that $\delta \neq -1$, since the value
$\delta=-1$ corresponds to infinite coupling constants
$\zeta=\pm\infty$. Therefore, Eq.\ (\ref{eq:d<0}) implies that
$A=0$ and leads to the condition
\begin{equation}
8\left( \delta +1\right) ^3\left[\left( \delta -1\right)\left[
\left( \delta -2\right) y^4-3\delta By^3\right] -\delta \left[
l^2C-2\left( \delta +1\right) B^2\right] y^2+\left( \delta
+2\right) Bl^2Cy\right] y^{2\delta }=0.
\end{equation}
It is easy to see that the coefficient of $y^{4+2\delta}$ never
vanishes. Consequently, no solutions exist for $\delta<0$.

The remaining case corresponds to the minimal coupling $\delta=0$,
in this case, Eq.\ (\ref{eq:Ein1}) implies that $\Psi={\rm
const.}$ and Eq.\ (\ref{eq:Ein2}) integrates directly to
\begin{equation}
F(r)=\frac{r^2}{l^2}-M\,,\quad \Psi ={\rm const.}
\label{eq:sBTZ}\, .
\end{equation}
This solution corresponds to the static BTZ solution \cite{BTZ},
which is once again no other than the $(2+1)$--anti--de~Sitter
space--time (see \cite{Einsenhart,Plebanski67,Weimberg,Garcia00}).

It is important to stress the fact that the three solutions
(\ref{eq:MZ}), (\ref{eq:d1}), and (\ref{eq:d2}) of the Einstein
equations (\ref{eq:Ein1})--(\ref{eq:Ein3}), also satisfy the
scalar field equation (\ref{eq:scalar}).

To conclude this letter let us summarize the results.

We have established the complete class of solutions within the
static circularly symmetric ansatz (\ref{eq:circular}) for the
structural function $F=F(r)$, and found that there only exist
solutions for the following values of the non--minimal coupling
constant $\zeta$:
\begin{itemize}

\item
For conformal coupling, $\zeta =1/8$, there exist two solutions,
the first one corresponds to the conformal black hole
(\ref{eq:MZ}) of Mart\'{\i}nez and Zanelli, and the second one is
a new conformal solution (\ref{eq:d1}), whose space--time
corresponds to the anti--de~Sitter space, with a non--trivial
scalar field.
\item
The non--minimal coupling $\zeta=1/6$ leads to a new solution
(\ref{eq:d2}), which space--time geometry is once again reducible
to the anti--de~Sitter form endowed with a non--trivial scalar
field.
\item
For minimal coupling $\zeta=0$, the scalar field reduces to a
constant, and the resulting space--time corresponds to the
$(2+1)$--anti--de~Sitter space--time (\ref{eq:sBTZ}).
\end{itemize}

The possibility of obtaining more general static, circularly
symmetric solutions, with two structural functions is still open.
The corresponding field equations are much more involved and hard
to integrate in general. Nevertheless, efforts have to be
undertaken to overcome the difficulties in the integration process
of this more general dynamical system.

\acknowledgments

We thank Friedrich W. Hehl for useful discussions and literature
hints. This research was partially supported by CONACyT Grants
32138E and 28339E, and by FOMES Grant: P/FOMES 98--35--15.

\end{document}